\documentstyle[aas2pp4]{article}
\begin{document}

\title{Observational Evidence for Superfluidity, and
Pinning, in the Core of Neutron Stars}
\author{M. Jahan-Miri} 
\affil{Institute for Advanced Studies in Basic Sciences, Zanjan
45195, IRAN}
\authoremail{jahan@iasbs.ac.ir}

\begin{abstract}
The observed large rates of spinning down after glitches
in some radio pulsars has been previously explained in terms of a
long-term spin-up behaviour of a superfluid
part of the crust of neutron stars. We argue that the suggested
mechanism is not viable; being
inconsistent with the basic requirements for a superfluid spin-up,
in addition to its quantitative disagreement with the data.
One may therefore conclude, for the first time, that the presence and the pinned
nature of neutron superfluid in the core of neutron stars is
evidenced by the existing observational data.
\end{abstract}

\section{Introduction}
The observational data we are invoking is more than a
decade old. Also, its implication is rather straightforward,
once it is realized that there exist no alternative
explanation for the data, based on the role of the
(superfluid in the) crust alone. The present discussion
tends to show the falacy in the only alternative possibilty
which has been raised to the opposite; thus paving the way
to the obvious, though so far abondoned, 
implication of the data on the large post-glitch excess
spin-down rates of radio pulsars for the role of the stellar core. 
The alternative suggestion has invoked a spinning up of the
crustal superfluid by the spinning down crust (``the container'')
over a time much larger than the associated relaxation timescale
(Alpar, Pines \& Cheng 1990; hereafter APC90). However, a closer
look at the relative rotation of the superfluid
and its vortices reveals that the suggested mechanism fails
quantitatively by, at least, more than an order of
magnitudes. In addition, the suggested spin-up
process is also argued to be in contradiction with the
well-known requirements for a superfluid spin-up. This will
lead us to the conclusion that the observed effect indicates 
the presence of a superfluid component, as well as the
pinning of its vortices, in the core of neutron stars.
We wish to emphasize that the point of the present discussion
is {\it only} to point at the above implication of the data, 
for which no valid observational evidence has been, so far,
invoked (eg., Alpar 1992). The conclusion of the paper would naturally
call for, and would be further supported by, a detailed and
successful modelling of the post-glitch response 
of a pinned superfluid component of the core.
The latter is nevertheless a separate and independent problem which
is not the subject of the present discussion, and which 
has been addressed previously (eg., Muslimov \& Tsygan 1985;
Chau, Cheng \& Ding 1992; Sedrakian \& Sedrakian 1995; Jahan-Miri 1996). 

\section{Overview}
The spin-down rate $\dot\Omega_{\rm c}$ of the crust of a neutron
star, with a moment of inertia $I_{\rm c}$, obeys (Baym et.~al. 1969a)
\begin{eqnarray}
 I_{\rm c} \dot\Omega_{\rm c} = N_{\rm em} -
                \Sigma I_{\rm i}\dot\Omega_{\rm i} 
\end{eqnarray}
where $N_{\rm em}$ is the negative electromagentic torque on
the star, and 
$\dot\Omega_{\rm i}$ and $I_{\rm i}$ denote the rate of change of
rotation frequency and the moment of inertia of each of the separate
components, respectively, which are summed over. Steady state
implies \(\dot\Omega_{\rm i} =\dot\Omega_{\rm c} = \dot \Omega \equiv
{N_{\rm em} \over I}\), for all $i$, where \( I= I_{\rm c}+\Sigma I_{\rm i}\)
is the total moment of inertia of the star. 
Different models for the post-glitch recovery, and 
in particular the model of vortex-creep, aim to explain
it in terms of a decoupling-recoupling of the superfluid
component in the {\em crust} of neutron stars
(Alpar et.~al. 1984; Jones 1991; Epstein et.~al. 1992). The
rest of the star including the core (superfluid) is assumed in
these models to be rotationally coupled to 
the non-superfluid constituents of the crust, on
timescales much shorter than that resolved in a glitch.

The role of a superfluid component of a neutron star in its
post-glitch behavior is understood as follows. 
Spinning down (up) of a superfluid at a given rate is associated with a
corresponding rate of outward (inward) radial motion of its vortices.
If vortices are subject to pinning, as is assumed for the superfluid in the
crust of a neutron star, a spin-down (up) would require unpinning of the
vortices. This may be achieved under the influence of a Magnus force
$\vec{F}_{\rm M}$ acting on the vortices, which is given, per unit length,
as (eg., Sauls 1989) 
\begin{eqnarray}
\vec{F}_{\rm M} & = & -  \rho_{\rm s} \vec{\kappa} \times
                                (\vec{v}_{\rm s} - \vec{v}_{\rm L})
\end{eqnarray}
where $\rho_{\rm s}$ is the superfluid density, $\vec{\kappa}$ is the
vorticity of the vortex line directed along the rotation axis (its
magnitude $\kappa = { h \over 2 m_{\rm n}}$ for the neutron superfluid,
where $m_{\rm n}$ is the mass of a neutron), and $\vec{v_{\rm s}}$ and
$\vec{v_{\rm L}}$ are the local superfluid and vortex-line velocities.
Thus, if a lag \( \omega \equiv \Omega_{\rm s} - \Omega_{\rm c}\) exist
between the rotation frequency $\Omega_{\rm s}$ of the superfluid and that
of the vortices (pinned and co-rotating with the crust) a radially
directed Magnus force \((F_{\rm M})_r= \rho_{\rm s} \kappa r \omega \)
would act on the vortices, where $r$ is the distance from the 
rotation axis, and \(\omega >0 \) corresponds to an outward directed
$(F_{\rm M})_r$, vice-versa. The crust-superfluid may
therefore follow the steady-state spin-down of the star by maintaining
a critical lag $\omega_{\rm crit}$ which will enable the vortices to
overcome the pinning barriers. The critical lag is thus defined through
the balancing $(F_{\rm M})_r$ with the pinning forces.

At a glitch a sudden
increase in $\Omega_{\rm c}$ would result in \( \omega<\omega_{\rm crit}\), 
hence the superfluid becomes decoupled and could no longer follow the
spinning down of the star
(ie. its container). If, as is suggested (Anderson \& Itoh 1975), the
glitch is due to a sudden outward release of some of the pinned vortices 
the associated decrease in $\Omega_{\rm s}$ would also add to the
decreas in $\omega$, in the same regions. Therefore a fractional
increase ${\Delta \dot \Omega_{\rm c} \over \dot \Omega_{\rm c}}$ 
same as the fractional moment
of inertia of the decoupled superfluid would be expected. This situation
will however persist only till \( \omega=\omega_{\rm crit}\) is restored
again (due to the spinning down of the crust) and the superfluid recouples, as
illustrated by $\Omega_{\rm s}$ and $\Omega_{\rm c}$
curves in Fig.~1. The vortex creep model suggests (Alpar et. al. 1984)
further that a superfluid spin-down may be achieved even while
\( \omega < \omega_{\rm crit}\), due to the creeping of the vortices
via their thermally activated and/or quantum tunneling motions. A
superfluid spin-down with a 
steady-state value of $\omega < \omega_{\rm crit}$, and also a
post-glitch smooth gradual turn over to the complete recoupling for each 
superfluid layer is thus predicted in this model.
Notice however that due to the varying amplitude of the glitch-induced jump
in $\Omega_{\rm s}$  (Alpar et. al. 1984) the sharp recoupling of each 
layer, in the absence of creeping, may occur at a different time. 
Hence the post-glitch recovery of $\dot\Omega_{\rm c}$, being accomplished
over an extended time period as the different layers of the superfluid
recouple, could behave
smoothly even in the absence of any creep. The induced
$\Delta\dot\Omega_{\rm c} \over \dot\Omega_{\rm c}$ during a superfluid
decoupling phase according to the creep model is however the same as
(or slightly smaller than) otherwise.

\section{The Problem}
Recent glitches in Vela, and one in PSR~0355+54, have shown values of
${\Delta \dot \Omega_{\rm c} \over
\dot \Omega_{\rm c}} > 10 \%$ with recovery timescales
$\sim 0.4$~d, and $\sim 44$~d, respectively
(Lyne 1987; APC90; Flanagan 1995).
The data hence imply (Eq.~1) that a part of the star with a fractional
moment of inertia $> 10\%$ (-- $60\%$) is decoupled from
the crust during its observed post-glitch
response. This is, however, in sharp contradiction with the above 
glitch models, since for the moment of inertia $I_{\rm crust}$
of the crustal superfluid \({I_{\rm crust} \over I} \lesssim 2.5\% \)
(APC90; Chau et.~al. 1993). It should be realized that the 
disagreement with the data is a fundamental shortcoming for
the crustal models; not just a quantitative mismatch. 
Because, the predicted increase for 
${\Delta \dot \Omega_{\rm c} \over \dot \Omega_{\rm c}}$
in these models is naturally bound to be smaller than the
fractional moment of inertia of the superfluid (also see the
best fit results of
Alpar et.~al. 1993; Chau et.~al. 1993); except for the
possibility raised in APC90 which is the point of issue in
the following.
                                       
APC90 suggest that the observed large values of
${{\Delta \dot \Omega_{\rm c} \over \dot \Omega_{\rm c}}}
\sim 20 \%$
over a time $\sim 0.4$~d, in Vela, could be 
accounted for by assuming that a part of the crustal superfluid
{\it spins up}, over the {\it same} period of time.
The superfluid would then act as a source of an
additional spin-down torque on the rest of the star and  
could, in principle, 
result in ${\Delta \dot \Omega_{\rm c} \over \dot \Omega_{\rm c}}$
values much larger than its own fractional moment of inertia.
For this to be realized, a region of the crust-superfluid 
with a moment of inertia
$I_{\rm sp}$ and a spin frequency $\Omega_{\rm sp}$ has
been assumed to support a tiny (positive) steady-state lag 
$\omega_{\rm sp} \equiv \Omega_{\rm sp}
-\Omega_{\rm c} \sim 3.5 \times 10^{-6} \ {\rm rad \ s}^{-1}$, 
in contrast to the much larger steady-state value of the lag 
\( \omega \geq 10^{-2} \ {\rm rad \ s}^{-1} \) in the rest of
the crust-superfluid.
Hence, a glitch of a size \(\Delta\Omega_{\rm c} \sim 10^{-4}
\ {\rm rad \ s}^{-1}\) would result in a ``reversed''
situation where \( \Omega_{\rm c} >> \Omega_{\rm sp}\), and 
which is further suggested (APC90) to be followed by a spinning up
of the superfluid over a period $\tau_{\rm sp}$, as
indicated in Fig.~1a.

\section{Quantitative Check}
The above spin-up scenario is however unable to account for
the observed effect, quantitatively.
In contrast to the assumption of APC90 that the total original
frequency difference ($\Omega_{\rm c}-\Omega_{\rm sp}$), induced
by the glitch, is slowly relaxed during the period $\tau_{\rm sp}$
we argue that only a small fraction of it might be at all preserved
for any such ``long-term'' spin-up process. This is because
the superfluid would be otherwise  rotating
much slower than its vortices which are, by virtue of their assumed
pinning, co-rotating with the
crust (see Fig.~1a); that is, the superfluid rotational lag with its
vortices would be {\em much larger} than the associated critical
lag. If so, the pinning could not impede the vortex motions (\S~2) and
a fast superfluid spin-up should be expected. In order to allow for
the suggested
large frequency difference between the superfluid and the crust,
and at the same time for a rotational lag (between the superfluid and
its vortices) {\em smaller} than the associated critical value one
needs to further assume
that the pinning is ``switched off'', in contradiction to the 
assumed pinning conditions. Nevertheless, the superfluid relaxation
would be still expected to take place very fast, as is further discussed
below. It may be recalled that the
critical lag is, by definition, the minimum lag required
for the Magnus force on vortices to overcome the pinning forces.
When the instantaneous lag exceeds its critical value, then the pinning
forces (in the azimuthal direction) would act further as a major source
for the torque on the superfluid, resulting in a relaxation even
faster than in the absence of any pinning (see, eg.,
Tsakadze \& Tsakadze 1980; Adams, Cieplak \& Glaberson 1985). 

The above argument may be further clarified by directly considering
the behaviour of the rotation frequency $\Omega_{\rm L}$ of
the {\it vortices} (in the region with $\Omega_{\rm sp}$) 
at the glitch, which has not been explicitly specified in APC90.
We, therefore, discuss the two exclusive possibilities that might
arise, and could be, in principle, physically justified.
Both cases lead, however, to the same conclusion; intermediate
cases for which no justification exist should naturally fall
in between (and there is no indication in APC90 for any special
effect due to such cases). It may be noted that the cases to be considered 
should not be, however, paralleled to the classification of
(strong, weak, superweak) pinning regions, as invoked in the literature
on the vortex creep model. The latter is based on the relative magnitude
of the critical lag and reflects the long term behaviour of the superfluid
relaxation toward its steady state lag value. In contrast, the following
two cases concerns the instantaneous response of the originally pinned
vortices
upon a sudden jump in the rotation rate of the container (see Fig.~1).
Either, {\bf a)} the vortices are spun up along with the crust (container),
that is they remain pinned {\em during} the sudden spin-up of the container. 
Or else {\bf b)} if their pinning is temporarily broken, they must relax to 
a state of co-rotating with the superfluid itself. Hence, either
$\Omega_{\rm L} = \Omega_{\rm c}$ (pinning conditions),
or else $\Omega_{\rm L} = \Omega_{\rm sp}$ (Hemholtz theorem),
just {\it at} the beginning of the interval
$\tau_{\rm sp}$ after the jump in $\Omega_{\rm c}$
(see Fig.~1a).

In the case {\bf (a)}, the superfluid must
have been also spun up, along with
the crust and the vortices, to (at least) a state such that
\( \Omega_{\rm sp}-\Omega_{\rm c}= -\omega_{\rm sp}\)
(contrast with APC90's picture as depicted in Fig.~1a);
otherwise the pinning would be broken (contrary to the
assumption) by the associated radial Magnus force on the
vortices. Experiments on superfluid Helium have indeed showed
(Tsakadze \& Tsakadze 1975; Alpar 1978; Tsakadze \& Tsakadze 1980)
that a pinned superfluid either is spun
up along with its vessel, or it never does so during the subsequent
spinning down of the vessel (when
$|\Omega_{\rm sp} -\Omega_{\rm L}| < \omega_{\rm sp}$).
Also according to the vortex creep model (see Eq.~6 in Cheng
et.~al. 1988), the superfluid spin-up (-down) timescale
should be extremely short ($\lesssim 10^{-7} \ {\rm s}$ for the
above case in Vela) given that  
$|\Omega_{\rm sp} -\Omega_{\rm L}| \geq \omega_{\rm sp}$,
as would be the case if only the vortices, but not the
superfluid, are assumed to be spun up with the crust. 

The case {\bf (b)}, on the other hand,
is not in accord with the general pinning conditions assumed in
the vortex creep model, and is not likely to be invoked
in that context. Nevertheless, the superfluid spin-up in the
crust of a neutron star, for such a case of free unpinned
vortices, is again expected to occur over very short
timescales. The longest timescale for the spin-up
of the crust by freely moving vortices, due to nuclear
scattering alone, has been estimated (Epstein et.~al. 1992) to be only 
$\lesssim 5$~s, for the Vela pulsar. Such a fast spin-up (-down)
of the superfluid in the crust of a neutron star, by the
freely moving vortices, is in fact invoked in the vortex
creep model as the cause of glitch itself (Anderson \& Itoh
1975). Therefore, and in either cases ({\bf a} or {\bf b}),
the superfluid would be spun up to (at least) a frequency
such that \(\Omega_{\rm sp}-\Omega_{\rm c}=-\omega_{\rm sp}\), 
within a period of only few seconds while the jump in
$\Omega_{\rm c}$ takes place at the glitch (see Fig.~1b). 
Notice that the steady-state lag $\omega_{\rm sp}$
according to the vortex creep model would be slightly smaller
than the critical lag (though slightly larger in the absence
of any creeping).
The difference is however a tiny fraction of the critical value
itself (Alpar et. al.~ 1984), and is therefore neglected in the 
present discussion.

Hence, the
frequency difference $\Delta\Omega$ between the crust and
the superfluid, at the start of $\tau_{\rm sp}$, would be 
$\Delta\Omega = \omega_{\rm sp} \sim 3.5\times10^{-6}\
{\rm rad~s}^{-1}$. This is the total available range of
rotation frequency that one might, in principle, assume to be 
equilibrated any further between the crust and the superfluid. 
In contrast, a value of 
$\Delta\Omega = \Delta\Omega_{\rm c}\sim 1.3\times10^{-4}\
{\rm rad~s}^{-1}$ has been used in APC90. 
Assuming, for the time, that any further spin-up of the
superfluid may occur, its duration, $\tau_{\rm sp}$, may
be estimated from (Baym et.~al. 1969a; see also Eq.~2b in APC90) 
\begin{eqnarray}
       \left(\Delta\dot\Omega_{\rm c}\right)_{\rm sp}=
                {I_{\rm sp}\over I} { \Delta\Omega \over \tau_{\rm sp}}
\end{eqnarray}
where $(\Delta\dot\Omega_{\rm c})_{\rm sp}$ is the magnitude of
the change in $\dot\Omega_{\rm c}$ due to the spinning up of the
superfluid. Adopting the same parameter values as in APC90, 
ie. 
\({{(\Delta\dot\Omega_{\rm c})_{\rm sp}}/{\dot\Omega_{\rm c}}}=
0.2\), \(\frac{I_{\rm sp}}{I}=5.3\times10^{-3}\), \(\Delta\Omega=
\omega_{\rm sp}=
3.5\times10^{-6}\ {\rm rad~s}^{-1}\), and
$\dot\Omega_{\rm c}= 9.5\times10^{-11}\ {\rm rad~s}^{-2}$ for Vela, 
one derives from Eq.~3 a value of
\begin{eqnarray}
\tau_{\rm sp}< 0.3 \ {\rm hr}, 
\end{eqnarray}
which is {\it too short} as compared with the observed 
timescales $\sim 0.4$~d.
Thus, the crust-superfluid cannot be the cause for the observed
large spin-down rates even in the case of 1988 glitch of the
Vela pulsar, addressed in APC90. The disagreement
between the predicted and observed timesacles would be
even worse for the case of
1991 glitch of the same pulsar, having observed values of  
\({\Delta\dot\Omega_{\rm c}/{\dot\Omega_{\rm c}}} \sim 60\% \) 
over a similar relaxation time (Flanagan 1995). Also, 
an attempt to apply the same crust-superfluid spin-up scenario
to the case of PSR~0355+54 would result in more than
{\it three} orders of magnitudes difference between the
predicted $\tau_{\rm sp}$ and the observed relaxation time
$\sim 40$~days.

\section{Superfluid Spin-up}
Moreover, the suggested spin-up scenario of APC90
should be dismissed at once
since the required torque on the superfluid, during
$\tau_{\rm sp}$, may not be realized at all, under the assumed
conditions (see Fig.~1a). That is
the pinned superfluid could not be spun {\em up} by
the crust (ie. its container) while the latter is
spinning {\em down}.
This is simply because a spinning down vessel (or even one with
a stationary constant rotation rate) albeit rotating faster than
its contained superfluid could not result in any further {\em spin
up} of the vortex lattice which is, by virtue of the assumed
pinning, already {\em co-rotating} with it!. As is
well-known, an inward radial motion of the vortices,
associated with a spin-up of the superfluid, may only
occur in the presence of a corresponding forward azimuthal 
external force acting on them. This is indeed a trivial
fact, considering that any torque on the bulk superfluid may
be applied only by means of the
vortices. However, no forward azimuthal force
(via scattering processes between the constituents particles
of the vortex-cores and the crust) may be exerted by the
spinning down crust on the vortex lattice which is already
co-rotating with it. Thus, no superfluid spin-up might be
expected to occur during the interval $\tau_{\rm sp}$,
namely after the spinning up of the crust  and 
{\it vortices} has been accomplished (compare Fig.~1a with
Fig.~1b). Notice that the requirement for a superfluid spin-up,
namely that the vortices be rotating slower than the crust while
the latter is itself slowing down, is inherently inconsistent
with the assumed pinning conditions.

The suggested long-term spin-up in APC90 is a generalization of
the vortex creep model for the case of a {\em negative}
lag, in contrast to the usual post-glitch spinning-down due
to a positive lag invoked in the model (eg., Pines \& Alpar 1992).
However, according to the
existing formulation of the vortex creep model, a  
spinning up of the superfluid would require a {\it positive} 
torque $N_{\rm em}$ acting on the whole star
(see, eg., Eqs.~28, and 38 in Alpar et.~al 1984). 
Application of the same formalism to
the suggested spin-up episode in presence of the given 
{\it negative} $N_{\rm em}$ (which is attempted in Eq.~5 of APC90)
is not, a priori, justified; it is contradictory.

The vortex creep model suggests that a radial Magnus force, due to
a superfluid rotational lag, results
in a radial bias in the otherwise randomly directed
creeping of the vortices (Alpar et. al.~ 1984).
This might be, mistakenly, interpreted to require that given a negative
lag the inward creeping motion of the vortice, hence a superfluid spin-up,
should necessarily follow, irrespective of the presence or absence of
the needed torque on the superfluid. It should be however noted that 
the role of driving the vortices inward, ie. spinning up
of the superfluid, may not be assigned to the Magnus force. The 
Magnus force associated with the rotational lag is a
{\em radial} force and is also exerted by the superfluid {\em itself};
both properties disqualifying it for being the source of a torque. 
Thus, the point which is to be emphasized here is that
the obvious requirement for a spin-up process, namely the corresponding
torque, is indeed missing in the suggested mechanism of APC90.  
Accordingly, the inability of the superfluid to be spun up, in this
case, should {\it not} be understood as 
a direct consequence of the pinning preventing a vortex radial motion,
which could be possibly cured by any assumed radial creeping motions.
Rather, the vortices under the assumed conditions do not have any
``tendency'' for an inward radial motion, in spite of the
presence of an inward radial Magnus force which is balanced by the
pinning forces. A change in the spin frequency of a superfluid involves
not only a radial motion of the vortices but also a corresponding
azimuthal one, as may be also verified from the solution of equation of
motion of vortices during a superfluid rotational relaxation
(see Eq.~9 in Alpar \& Sauls 1988 and Eq.~4 in Jahan-Miri 1998).
The torque may be transmitted during only such an azimuthal motion
and would nevertheless require and initiate a radial motion as well. 
Therefore, purely {\it radial} creeping of the vortices, which
is implied by the existing formulation of the vortex creep model,
may not be invoked as a spinning-up mechanism during the transition
from \(\Omega_{\rm sp} - \Omega_{\rm L}= -\omega_{\rm sp}\) to
\(\Omega_{\rm sp} \gtrsim \Omega_{\rm L}\).
The superfluid would rather
remain decoupled at a constant value of $\Omega_{\rm sp}$
(if not spinning down) during this transition which is achieved
due only to the spinning down of the crust, as illustrated in Fig.~1b.

\section{Concluding}
Decoupling of (a part of) the moment of inertia of the crust at a glitch,
from the rest of the corotating star, could readily account for the excess
post-glitch spin-down rates comparable to the fractional moment of inertia
of the decoupled part. The same preliminary fact applies to a (partial)
decoupling of a (pinned) superfluid component in the crust, or elsewhere in
the star, for that matter, as well. As suggested by APC90, a 
component (say, in the crust) may result in an even larger excess spin-down
rate for the star if the decoupled component is further assumed to be
spinning up while the rest
of the star is spinning down; ie. a {\em negative} coupling instead of
a mere decoupling. Nevertheless, as we have shown the suggested mechanism
of APC90 
for such a negative coupling of a pinned superfluid part in the crust
of a neutron star not only fails quantitatively to account for the observed
effect in pulsars, it is also ruled out conceptually since the required
torque on the superfluid could not be realized at all.

Thus, the (pinned) crust-superfluid may not be invoked as
the cause of the observed large changes in the spin-down rates of
neutron stars. One is therefore left to conclude definitely that the
observed effect, over timescales of a day and more, is
caused by the stellar core, having a large moment of inertia
$I_{\rm core}$, with \( {I_{\rm core} \over I} \sim 90~\%$. This is, 
however, by itself an obvious
observational evidence for the {\em superfluidity} and at the same time
{\em pinning} of the vortices in the core of a neutron star, given the
current ``standard" picture of the interior of these stars
(Sauls 1989; Pines \& Alpar 1992). For, a non-superfluid core would
couple to the crust on very short timescales ($< 10^{-11}$~s)
(Baym et.~al 1996b), and could not have any footprint left in the
observed post-glitch relaxation. Also, a core-superfluid with free
(unpinned) vortices would be again expected to have very short coupling
timescales of the order of less than one or two minutes
(Alpar \& Sauls 1988; Pines \& Alpar 1992; Jahan-Miri 1998).

The above conclusion, for the presence of pinned core superfluid,
being a requirement of the observational data is obviously independent
of any specific model for the pinning in the core and the post-glitch
effects of the superfluid relaxation. It is appreciated that
if the core-superfluid is assumed to be subject to a pinning as a
{\em whole} and also to be {\em completely} decoupled at a glitch then
the induced post-glitch values of
${\Delta \dot \Omega_{\rm c} \over \dot \Omega_{\rm c}}$ would 
be much larger than what has been usually observed for the glitching pulsars.
Nevertheless, there is no requirement for the pinning of the core
superfluid that it has to comply to such tight restrictions. Given the large moment
of inertia of the core, a pinning of some parts and/or a partial
decoupling of the pinned component may as well be invoked so that it
would result in post-glitch spin-down rates similar to that observed.
Indeed, a pinning between the vortices and the proton fluxoids 
(Muslimov \& Tsygan 1985; Sauls 1989)
might offer a resolution, as is indicated by our preliminary
results (Jahan-Miri 1996; Jahan-Miri 1999), even if all of the vortices
in the core are assumed to be involved. The novel feature
of this pinning mechanism, as it affects the rotational dynamics of the
core superfluid, is the {\em moving} nature of
the {\em pinning sites} which might allow for a spinning down of the
superfluid even while the vortices are kept pinned! Also, the model
of ``proton-vortex cluster'' (Sedrakian \& Sedrakian 1995), invoking a
somewhat different picture of the pinning in the core of neutron stars,
might serve as another candidate mechanism.
In any case, while the effects due to the pinned core-superfluid  
on the observable post-glitch relaxation of radio pulsars needs to be further
demonstrated, the observed large excess spin-down rates is pointing at  
an apparent role played by the stellar core, for which it ought to be in
a pinned superfluid state.

Raman Research Institute is acknowledged for kindly providing
financial support as well as the research  
environment where this work could be initiated.

\begin{figure}
\caption{As is shown, schematically, by the {\it thin} and
the {\it dotted} lines, a superfluid, as in the crust of a neutron star,
whose vortices are pinned and co-rotating, at
a rate $\Omega_{\rm c}$, with its spinning down vessel,
may spin-down at the same rate as the vessel provided
a certain lag between its rotation frequency
$\Omega_{\rm s}$ and $\Omega_{\rm c}$ is maintained. A decrease in
the lag, say at a glitch, results in a decoupling of the
superfluid
and hence a larger spin-down rate for the vessel. The superfluid
starts spinning down again once the lag is increased to its steady-state 
value. The {\it thick} line represents the 
superimposed behavior of a second superfluid
component, $\Omega_{\rm sp}$, which supports a much smaller
steady-state lag. Its behaviour as implied in APC90 is shown
in the panel {\bf (a)}, in contrast to that suggested here,
shown in {\bf (b)}.
The unresolved rising time of $\Omega_{\rm c}$ at the glitch
instant $t=t_{\rm g}$ should be
reckoned as a measure of the expected coupling time of the superfluid
due to free motion of vortices, in contrast to the suggested 
spin-up period $\tau_{\rm sp}$ in {\bf (a)}. }
\label{f2}
\end{figure}

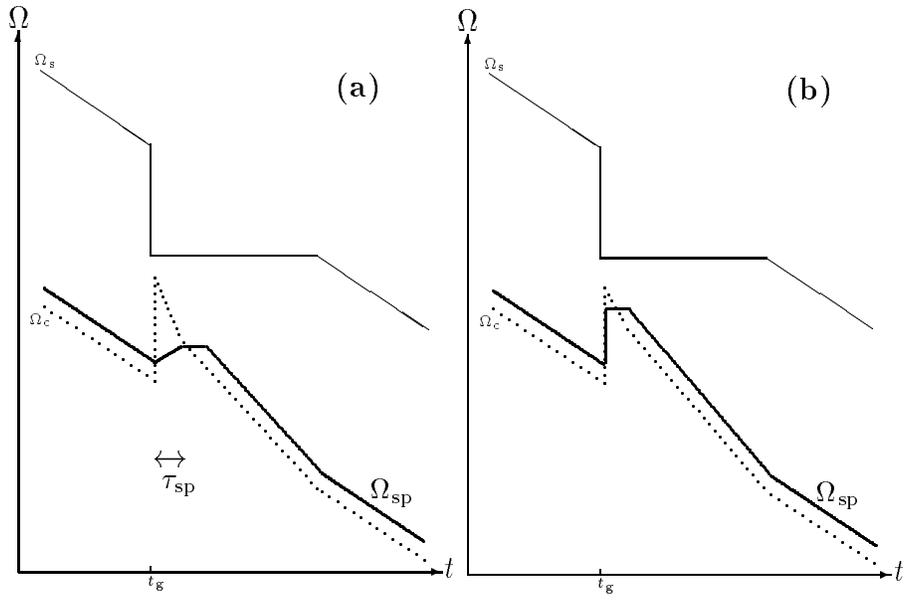
\end{document}